\documentclass[apj]{emulateapj}
\usepackage{apjfonts}
\usepackage[usenames,dvips]{color}

\begin{document}
\shorttitle{36 and 229~GHz Methanol Masers in DR21}
\shortauthors{Fish et al.}
\title{First Interferometric Images of the 36~GH\lowercase{z}
  Methanol Masers in the DR21 Complex}
\author{Vincent L.\ Fish\altaffilmark{1},
        Talitha C.\ Muehlbrad\altaffilmark{1,2},
        Preethi Pratap\altaffilmark{1,3},
        Lor\'{a}nt O.\ Sjouwerman\altaffilmark{4},
        Vladimir Strelnitski\altaffilmark{5},
        Ylva M.\ Pihlstr\"{o}m\altaffilmark{6,7},
        \&
        Tyler L.\ Bourke\altaffilmark{8}
}
\email{vfish@haystack.mit.edu}
\altaffiltext{1}{MIT Haystack Observatory, Route 40, Westford, MA 01886, USA}
\altaffiltext{2}{Current address: Texas Lutheran University, 1000 W.\ Court
  Street, Seguin, TX 78155, USA}
\altaffiltext{3}{Current contact: ppratap2002@gmail.com}
\altaffiltext{4}{National Radio Astronomy Observatory, P.O.\ Box O,
  Socorro, NM 87801, USA}
\altaffiltext{5}{Maria Mitchell Observatory, 4 Vestal Street,
  Nantucket, MA 02554, USA}
\altaffiltext{6}{Department of Physics and Astronomy, University of
 New Mexico, 800 Yale Boulevard NE, Albuquerque, NM 87131, USA}
\altaffiltext{7}{Adjunct astronomer at NRAO}
\altaffiltext{8}{Harvard-Smithsonian Center for Astrophysics, 60
  Garden Street, Cambridge, MA 02138, USA}
\begin{abstract}
Class~I methanol masers are believed to be produced in the
shock-excited environment around star-forming regions.  Many authors
have argued that the appearance of various subsets of class~I masers
may be indicative of specific evolutionary stages of star formation or
excitation conditions.  Until recently, however, no major
interferometer was capable of imaging the important 36~GHz transition.
We report on Expanded Very Large Array observations of the 36~GHz
methanol masers and Submillimeter Array observations of the 229~GHz
methanol masers in DR21(OH), DR21N, and DR21W.  The distribution of
36~GHz masers in the outflow of DR21(OH) is similar to that of the
other class~I methanol transitions, with numerous multitransition
spatial overlaps.  At the site of the main continuum source in
DR21(OH), class~I masers at 36 and 229~GHz are found in virtual
overlap with class~II 6.7~GHz masers.  To the south of the outflow,
the 36~GHz masers are scattered over a large region but usually do not
appear coincident with 44~GHz masers.  In DR21W we detect an
``S-curve'' signature in Stokes V that implies a large value of the
magnetic field strength if interpreted as due to Zeeman splitting,
suggesting either that class~I masers may exist at higher densities
than previously believed or that the direct Zeeman interpretation of
S-curve Stokes V profiles in class~I masers may be incorrect.  We find
a diverse variety of different maser phenomena in these sources,
suggestive of differing physical conditions among them.
\end{abstract}

\section{Introduction}

Methanol masers are often found in star-forming regions.  There are
two sets of transitions seen to produce methanol masers.  Class~I
methanol masers (most importantly the 36 and 44~GHz transitions) are
believed to be collisionally excited, while class~II masers (including
the 6.7 and 12~GHz transitions) are radiatively excited
\citep{cragg1992}.  Class~I and class~II methanol masers are sometimes
both found in association with the same source
\citep[e.g.,][]{slysh1994}, but the two classes of masers are very
rarely seen at the same velocity or in close (subarcsecond) spatial
overlap.

Class~I methanol masers, in which shocks dominate over infrared
radiation, have often been assumed to be tracing an earlier
evolutionary state of star formation than class~II methanol, water, or
OH masers \citep[e.g.,][]{ellingsen2006,breen2010}.  Subcategorization
of class~I masers by physical conditions may be possible
\citep{sobolev1993}, leading some authors to speculate that line
intensity ratios among the class~I masers may be a proxy for
evolutionary stage \citep{pratap2008}.  However, class~I maser studies
have traditionally been biased towards regions hosting other tracers
of star formation, and the cluster environments in which class~I
masers are found are usually quite complex, calling into question
traditional models of the evolutionary timeline of class~I masers
\citep[Section 4.4 of][and references therein]{voronkov2010-9.9}.

Furthermore, class~I masers have typically been observed with
single-dish telescopes, which can identify whether or not a particular
class~I transition produces masers in a region (and how bright they
are) but do not have the resolution to determine their location
relative to masers in other transitions.  Given the complex
environments associated with clustered star formation, high angular
resolution is required to identify the relations between masers and
excitation sources \citep[e.g.,][]{araya2009} and between multiple
transitions of methanol \citep{voronkov2006}.  Higher angular
resolution is also necessary to understand the physical conditions
that produce masers in each of the class~I transitions, which may not
be identical \citep[e.g.,][]{menten1991,johnston1992,sobolev1993}.

These concerns motivated \citet{pratap2008} to do an unbiased
single-dish search for class~I methanol masers in nearby molecular
clouds, resulting in the detection of new class~I maser features.
Several sites within these clouds host previously known 44~GHz
methanol masers, many of which have been mapped interferometrically
\citep[e.g.,][]{mehringer1997,kogan1998,kurtz2004}.  Imaging the
36~GHz masers, the other bright transition seen in numerous sources
\citep{haschick1989,berulis1990,liechti1996,pratap2008}, has
heretofore not been possible due to the lack of interferometers
operating at this frequency.  However, recent upgrades to the
Australia Telescope Compact Array and the Expanded Very Large Array
(EVLA) are allowing the first arcsecond-resolution images of 36~GHz
masers to be produced \citep{sarma2009,sjouwerman2010,voronkov2010}.
In this Letter, we report on the first EVLA maps of the 36~GHz masers
in the DR21 star-forming complex.

\section{Observations and Data Analysis}

The EVLA was used to observe the 36.169~GHz $4_{-1} \rightarrow 3_0~E$
line of methanol in DR21(OH), DR21W, and DR21N on 2010 May 26.  The
array consisted of the 20 telescopes outfitted with Ka-band receivers.
The EVLA was in its most compact (D) configuration, providing a
synthesized beamwidth of approximately $2\farcs1 \times 1\farcs4$.
All three sources were observed in dual circular polarization centered
on a fixed sky frequency of 36.1731~GHz and correlated with the new
WIDAR correlator.  The 4~MHz observing bandwidth was divided into 256
spectral channels, giving a velocity coverage of 33~km\,s$^{-1}$ with
a channel spacing of 0.13~km\,s$^{-1}$.  Conversion from sky frequency
to LSR velocity was performed with the assistance of the EVLA Online
Dopset
tool\footnote{\url{http://www.vla.nrao.edu/astro/guides/dopset/}}.
Total on-source observing time was $\sim 12$~min per source.  Typical
single-channel noise levels were $\lesssim 20$~mJy\,beam$^{-1}$ near
the center of the field

Data reduction was carried out using the NRAO Astronomical Image
Processing System (AIPS).  The flux scale was set using 3C48.  Complex
gain calibration was done using J2048+4310.  EVLA transition issues
precluded accurate bandpass calibration, although it could be
determined from calibrator data that baseline phases were flat across
the entire observing band and amplitudes were flat across about 75\%
of the observing band.  Thus, our main results, including estimates of
the positions of the detected masers, should not be significantly
affected, although bandpass effects may lead to amplitude errors of a
few percent over most of the observed velocity range.  Data were
imaged, and flux densities were corrected for primary beam
attenuation.

Minor velocity errors may be introduced by two effects.  First, the
uncertainty in the rest frequency of the observed line
\citep[$36\,169.265 \pm 0.030$~MHz;][]{muller2004} is equivalent to a
velocity uncertainty of 0.25~km\,s$^{-1}$ (2 channels).  Second,
observations were performed at a fixed sky frequency, since Doppler
tracking was unavailable.  The sky frequency associated with the rest
frequency drifted by about 0.015~MHz over the course of the
observations, possibly resulting in an insignificant spectral
broadening of maser features.  Since the net uncertainty in
determining the central velocity due to fixed-frequency observing ($<$
a few $\!\!\times 0.01$~km\,s$^{-1}$) is only a fraction of the
channel separation (and much smaller than a typical maser linewidth),
the data were not corrected with CVEL.

We present 229.758~GHz $8_{-1} \rightarrow 7_0~E$ maser and 226~GHz
continuum SMA data for all of our sources (P.\ Pratap et al., in
preparation).  The angular resolution of these observations is $\sim
1\farcs2 \times 0\farcs9$.  The bandwidth of the 229~GHz observations
covered the velocity range from $-50$ to $+81$~km\,s$^{-1}$ with a
spectral resolution of 0.13~km\,s$^{-1}$.  The flux densities of the
229~GHz masers were corrected for primary beam attenuation.  The rms
noise near field center was approximately 0.15~Jy\,beam$^{-1}$ for
DR21(OH) and DR21N and 0.3~Jy\,beam$^{-1}$ for DR21W.

When possible, we compare our data against three other class~I
methanol maser transitions from the literature: 44~GHz ($7_0
\rightarrow 6_1~A^+$), 84~GHz ($5_{-1} \rightarrow 4_0~E$), and 95~GHz
($8_0 \rightarrow 7_1~A^+$).  For DR21N, we also used archival 44~GHz
data taken with the VLA in B-configuration on 2006 Jun 16.

\section{Results}

\begin{figure*}
\begin{center}
\resizebox{0.65\hsize}{!}{\includegraphics{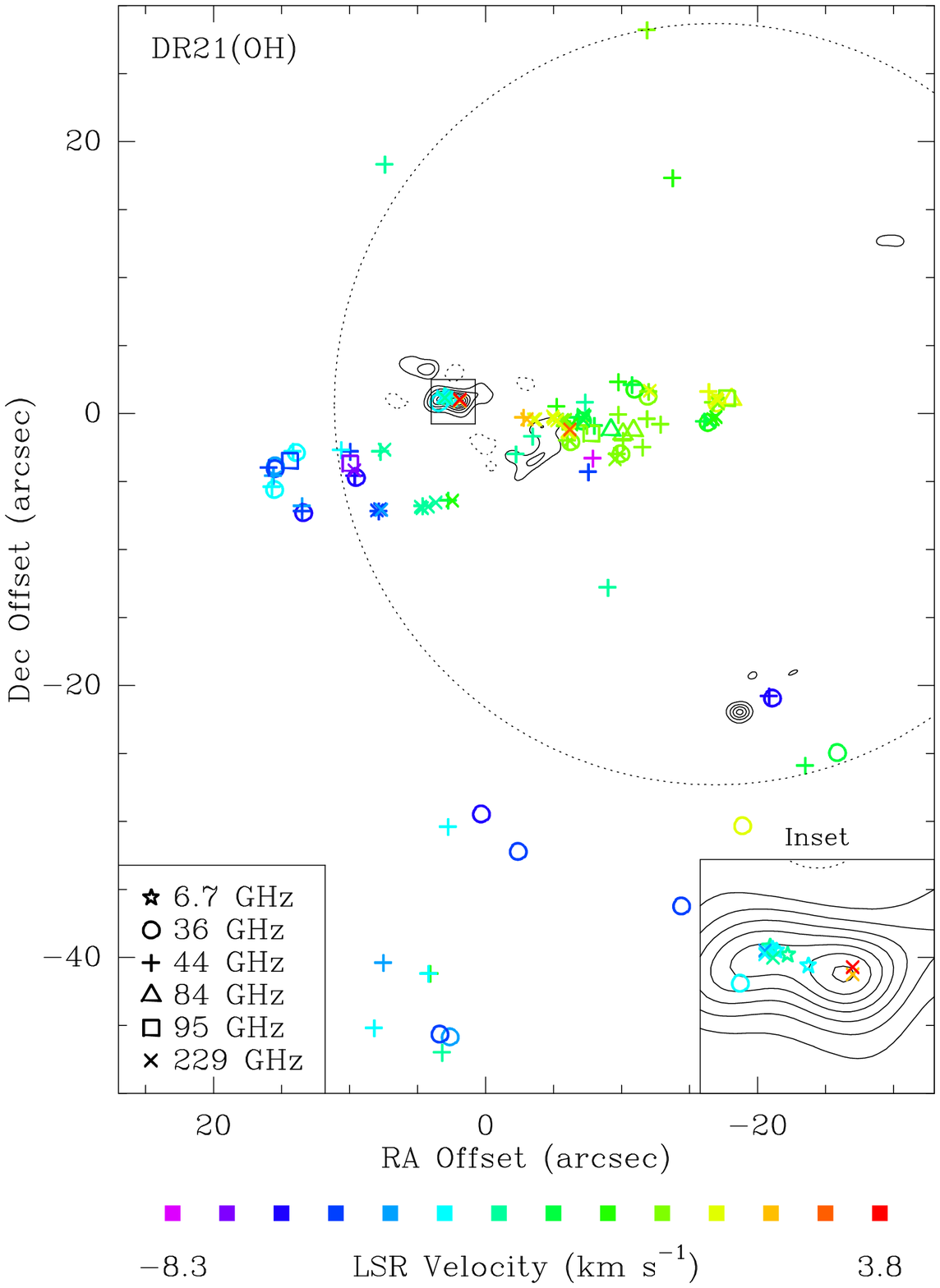}}
\end{center}
\caption{Class~I methanol maser plot of DR21(OH) showing the 36~GHz
  (this work), 44~GHz \citep{araya2009}, 84~GHz \citep{batrla1988},
  95~GHz \citep{plambeck1990}, and 229~GHz (P.\ Pratap et al., in
  preparation) class~I transitions and the 6.7~GHz class~II transition
  \citep{harvey-smith2008}.  Contours indicate 226~GHz continuum
  (P.\ Pratap et al., in preparation).  The angular resolution of the
  84 and 95~GHz data was lower than for the 36~GHz masers and should
  therefore be taken as indicative only of the general distribution
  and velocity structure of the emission in these transitions, which
  are consistent with nearby class~I masers in other transitions.  The
  dotted circle indicates the primary beam of the 226~GHz continuum
  and 229~GHz maser observations.  The inset box shows an enlargement
  of the $3\farcs25$ square box near the origin, where 6.7, 36, and
  229~GHz masers appear in close proximity and in the same velocity
  range.  Coordinates are relative to the EVLA pointing center
  $20^\mathrm{h}39^\mathrm{m}00\fs8, +42\degr22\arcmin48\farcs0$
  (J2000).
\label{fig-map-oh}
}
\end{figure*}

\begin{figure*}
\begin{center}
\resizebox{0.45\hsize}{!}{\includegraphics{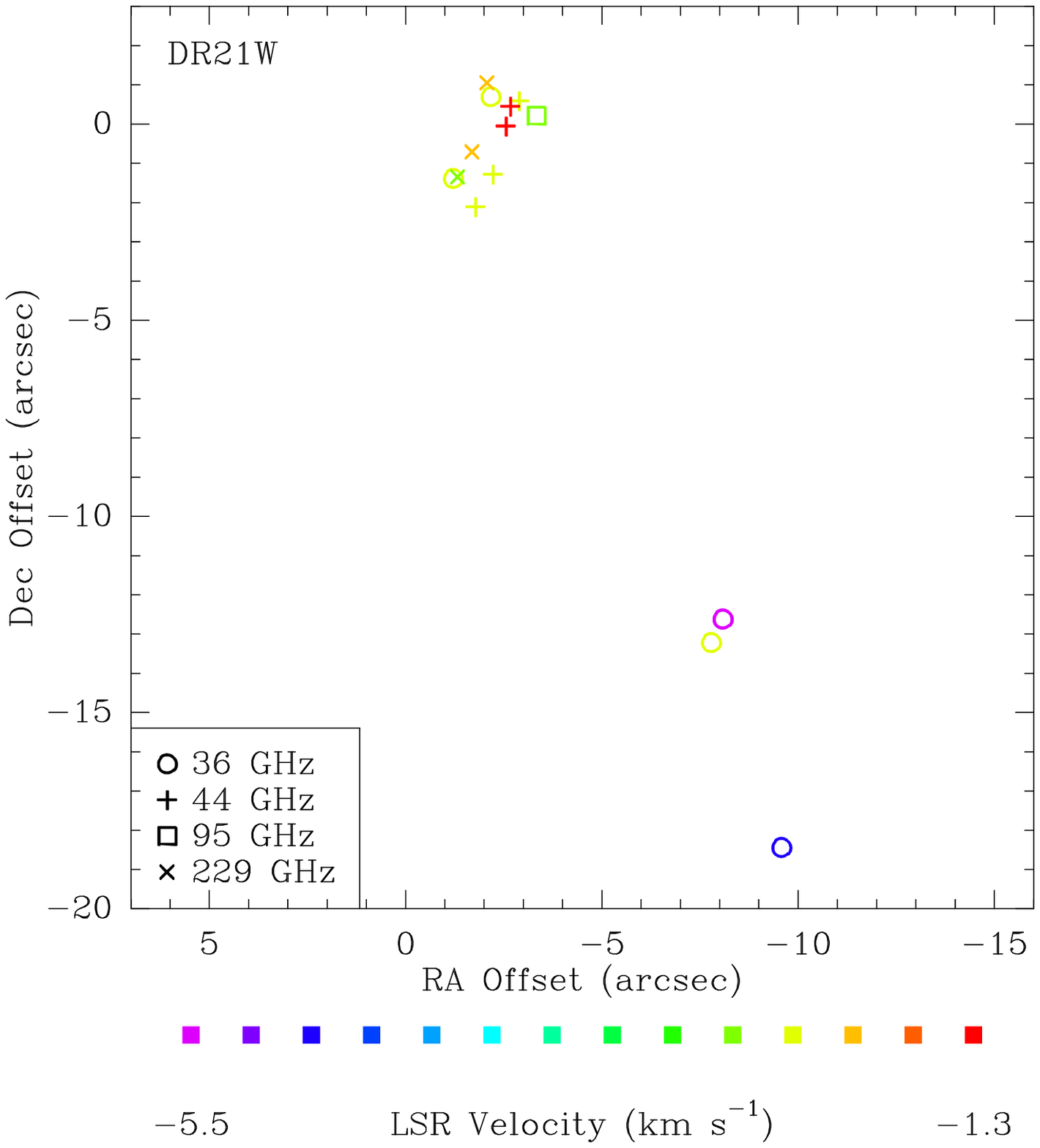}}
\resizebox{0.45\hsize}{!}{\includegraphics{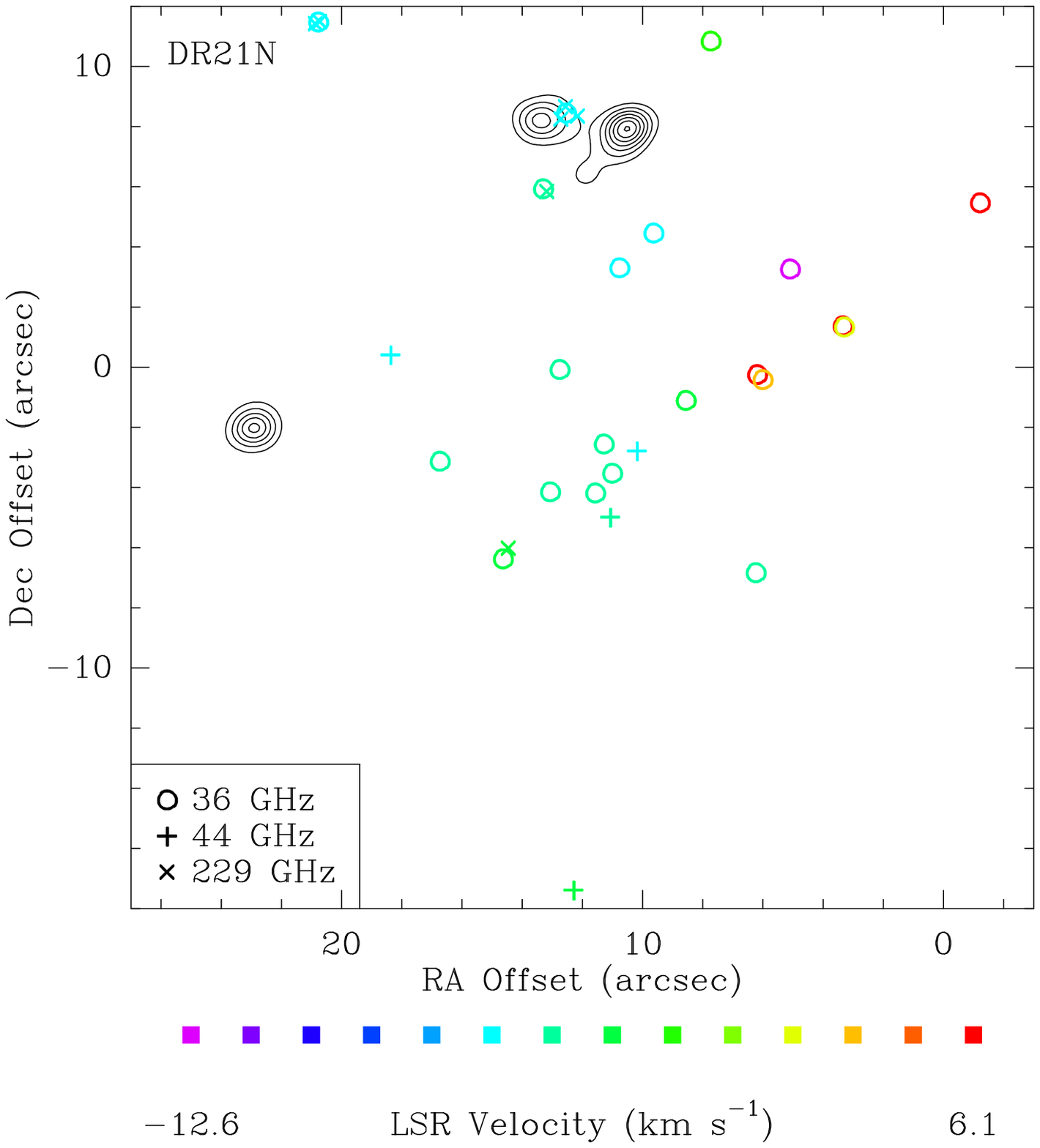}}
\end{center}
\caption{Class~I methanol maser plots of DR21W (left) and DR21N
  (right).  Symbols are as in Figure~\ref{fig-map-oh}.  The 44~GHz
  masers in DR21W are from \citet{kogan1998}.  The 44~GHz data in
  DR21N were taken with the VLA in B-configuration.  No 226~GHz
  continuum is detected in DR21W.  Every 229~GHz maser in DR21N is
  found in close proximity to a 36~GHz maser.  Coordinates are
  relative to the EVLA pointing center
  $20^\mathrm{h}38^\mathrm{m}55\fs0$, $+42\degr19\arcmin22\farcs0$ for
  DR21W and $20^\mathrm{h}39^\mathrm{m}02\fs0$,
  $+42\degr25\arcmin43\farcs0$ (J2000) for DR21N.
\label{fig-maps-n-w}
}
\end{figure*}

Parameters of detected masers are listed in Table~\ref{table-masers}.
We detect 49 36~GHz maser features in the three observed DR21 sources:
21 in DR21N, 23 in DR21(OH), and 5 in DR21W.  Figures~\ref{fig-map-oh}
and \ref{fig-maps-n-w} show the locations of the 36~GHz masers
relative to other class~I methanol transitions mapped with
interferometric resolution.  The LSR velocities of detected masers are
consistent with previous single-dish observations \citep{pratap2008}.

Spectra of the integrated flux of individual 36~GHz maser features are
often not well fit by a single Gaussian.  In some cases multiple
spectral peaks are evident, while in others there is evident skewness
in the spectral peak.  Both of these cases imply the existence of
structure on scales smaller than our synthesized beam size
(1--2\arcsec).  Similar effects are seen in the 44~GHz transition, in
which \citet{kurtz2004} detect 17 masers in DR21(OH) (several with
multiple spectral peaks) using 10 minutes of VLA D-configuration data,
while \citet{araya2009} detect 49 masers using a much longer
integration of C-configuration data.  When individual maser spectra in
our data do show a single clear peak, we find that typical linewidths
are on the order of 0.15 to 0.35~km\,s$^{-1}$.  There is also a
signature of very weak, extended methanol emission in some sources,
but the lack of proper bandpass calibration tempers our confidence in
this result.

There are 14 arcsecond-scale overlaps of a 36~GHz and a 229~GHz maser
at approximately the same velocity ($|v_{36} - v_{229}| \lesssim
0.4$~km\,s$^{-1}$).  Of these 14 associations, the 36~GHz maser is
brighter in 7, and the 229~GHz maser is brighter in the other 7,
indicating that either of the two $J_{-1} \rightarrow (J-1)_0~E$ lines
may be brighter (with the caveat that the 36 and 229~GHz observations
did not occur simultaneously and that maser variability is possible).
A similar phenomenon is seen in the $J_0 \rightarrow (J-1)_1~A^+$
series, where \citet{kalenskii1994} find that while the integrated
95~GHz flux is in general higher than that at 44~GHz, the integrated
fluxes of sources whose spectra consist of \emph{narrow-velocity} 44
and 95~GHz features (where the two transitions presumably produce
single masers in close spatial association) are comparable in the two
transitions.  Further high-resolution data will be necessary to
examine the relation between the flux densities of different maser
transitions seen in close spatial association with each other.

\begin{deluxetable}{lrrrr}
\tablecaption{Detected Masers\label{table-masers}}
\tablehead{
  \colhead{} &
  \colhead{RA} &
  \colhead{Decl.} &
  \colhead{} &
  \colhead{Peak} \\
  \colhead{} &
  \colhead{Offset} &
  \colhead{Offset} &
  \colhead{$v_\mathrm{LSR}$} &
  \colhead{Flux Density} \\
  \colhead{Source} &
  \colhead{(arcsec)} &
  \colhead{(arcsec)} &
  \colhead{(km\,s$^{-1}$)} &
  \colhead{(Jy)}
}
\startdata
DR21N    &  $-$1.21 &     5.46 &     6.07 &  0.96 \\
         &     3.30 &     1.33 &     0.76 & 23.72 \\
         &     3.35 &     1.36 &     5.42 &  0.17 \\
         &     5.10 &     3.25 & $-$12.58 &  0.31 \\
         &     6.01 &  $-$0.43 &     3.22 &  0.40 \\
         &     6.19 &  $-$0.26 &     6.07 &  0.19 \\
         &     6.24 &  $-$6.84 &  $-$3.38 &  1.63 \\
         &     7.73 &    10.83 &  $-$1.31 &  0.28 \\
         &     8.56 &  $-$1.12 &  $-$2.61 &  0.28 \\
         &     9.63 &     4.45 &  $-$4.94 &  0.25 \\
         &    10.77 &     3.29 &  $-$5.71 &  0.18 \\
         &    11.01 &  $-$3.54 &  $-$4.42 &  0.96 \\
         &    11.28 &  $-$2.57 &  $-$4.42 &  0.37 \\
         &    11.57 &  $-$4.19 &  $-$4.16 &  0.65 \\
         &    12.54 &     8.44 &  $-$5.71 &  0.15 \\
         &    12.75 &  $-$0.09 &  $-$4.42 &  0.31 \\
         &    13.07 &  $-$4.16 &  $-$4.03 &  0.60 \\
         &    13.30 &     5.92 &  $-$4.16 &  0.36 \\
         &    14.64 &  $-$6.39 &  $-$3.12 & 10.08 \\
         &    16.74 &  $-$3.14 &  $-$4.03 &  0.22 \\
         &    20.77 &    11.47 &  $-$5.20 &  6.46 \\
DR21(OH) & $-$25.86 & $-$24.96 &  $-$1.48 &  2.66 \\
         & $-$21.08 & $-$20.95 &  $-$6.01 &  0.36 \\
         & $-$18.88 & $-$30.34 &     0.47 &  2.67 \\
         & $-$16.99 &     0.75 &     0.47 & 30.26 \\
         & $-$16.34 &  $-$0.68 &  $-$1.35 &  1.26 \\
         & $-$14.39 & $-$36.24 &  $-$5.62 &  0.60 \\
         & $-$11.92 &     1.25 &  $-$0.05 &  1.31 \\
         & $-$10.96 &     1.77 &  $-$1.35 &  0.52 \\
         & $-$10.52 &  $-$1.72 &  $-$0.57 &  1.49 \\
         &  $-$9.93 &  $-$2.98 &     0.08 & 10.39 \\
         &  $-$7.10 &  $-$0.57 &  $-$1.61 &  2.39 \\
         &  $-$6.24 &  $-$2.10 &     0.08 &  1.81 \\
         &  $-$2.38 & $-$32.23 &  $-$5.49 &  0.82 \\
         &     0.33 & $-$29.47 &  $-$5.88 &  0.54 \\
         &     2.63 & $-$45.86 &  $-$4.20 &  7.61 \\
         &     3.38 & $-$45.66 &  $-$5.62 &  0.86 \\
         &     3.45\tablenotemark{a}&     0.78 &  $-$3.68 &  0.15 \\
         &     9.52 &  $-$4.73 &  $-$6.53 &  1.69 \\
         &    13.40 &  $-$7.31 &  $-$5.88 &  0.42 \\
         &    13.94 &  $-$2.89 &  $-$3.55 &  0.50 \\
         &    15.48 &  $-$3.87 &  $-$4.71 &  0.85 \\
         &    15.49 &  $-$4.07 &  $-$4.97 &  0.54 \\
         &    15.54 &  $-$5.61 &  $-$3.68 &  0.46 \\
DR21W    &  $-$9.57 & $-$18.45 &  $-$4.90 &  0.30 \\
         &  $-$8.08 & $-$12.63 &  $-$5.54 &  0.43 \\
         &  $-$7.78 & $-$13.23 &  $-$2.31 &  0.93 \\
         &  $-$2.16 &     0.69 &  $-$2.44 & 24.60 \\
         &  $-$1.21 &  $-$1.40 &  $-$2.44 & 82.29
\enddata
\tablenotetext{a}{Brightest channel of spectrally-broad weak emission;
  see Section \ref{dr21oh}.}
\tablecomments{Offsets are measured from
  $20^\mathrm{h}39^\mathrm{m}00\fs8$, $+42\degr22\arcmin48\farcs0$
  (J2000) for DR21(OH), $20^\mathrm{h}39^\mathrm{m}02\fs0$,
  $+42\degr25\arcmin43\farcs0$ for DR21N, and
  $20^\mathrm{h}38^\mathrm{m}55\fs0$, $+42\degr19\arcmin22\farcs0$ for
  DR21W.}
\end{deluxetable}

\subsection{DR21(OH)} \label{dr21oh}

The most striking feature of the masers in DR21(OH) is the central
region of the source, where numerous masers in all mapped transitions
occur in an approximately elliptical structure
(Figure~\ref{fig-map-oh}).  \citet{araya2009} identify more than 30
masers at 44~GHz associated with two shocks in a bipolar outflow.  To
within the positional uncertainty of our observations, all 14 of the
36~GHz masers in the outflow region are coincident with the
\citet{araya2009} 44~GHz masers.  The 84 and 95~GHz masers are all
located within the outflow region, although the lower angular
resolution of the observations (\citealp[$8\farcs3 \times
  6\farcs2$,][]{batrla1988}, and \citealp[$6\farcs5 \times
  4\farcs4$,][]{plambeck1990}, respectively) precludes more detailed
analysis of multitransition positional coincidences.  Many of the
229~GHz masers align with 36~GHz masers on the western side of the
outflow, but they appear to be more coincident with the 44~GHz masers
of \citet{araya2009}, especially in the eastern part of the outflow.

All transitions exhibit a redshift on the western side of the outflow
and a blueshift on the eastern side.  There is also a general velocity
gradient along the direction of the redshifted outflow to the west,
with LSR velocities generally increasing from the center outward.
Nevertheless, some of the masers on the northern and southern edges of
the outflow have velocities that do not fit this pattern.  The bright
36~GHz masers are located in the western side of the outflow, with the
brightest 36~GHz maser being located very near the brightest 44~GHz
\citep{araya2009} and 229~GHz masers.

Within the outflow, each detected 36~GHz maser is found in close
spatial association with a 44~GHz maser at approximately the same
velocity.  The reverse is not true; that is, there are numerous 44~GHz
features with no corresponding 36~GHz detection.  On the eastern side
of the outflow, 36~GHz masers are seen only at the eastern edge, while
other transitions are seen throughout.  In the central and western
portions of the outflow, many of the 44~GHz masers, especially those
that are highly redshifted or blueshifted compared to the velocity of
the largest portion of the masers in the western outflow, have no
accompanying 36~GHz features.

Class~II masers in the 6.7~GHz transition ($5_1 \rightarrow 6_0~A^+$)
only appear projected atop the bright continuum feature near the
origin in Figure~\ref{fig-map-oh} \citep{harvey-smith2008}.  Several
229~GHz class~I masers also appear, coincident with the 6.7~GHz masers
in some cases at the $0\farcs1$ level (one-tenth of the 229~GHz beam
size) but offset slightly to the east and south.  We also detect weak
36~GHz maser emission in the region, with spots ranging from 100 to
150~mJy detected in eight spectral channels between $-3.8$ and
$-2.6$~km\,s$^{-1}$, likely due to spectral blending of maser features
that are not fully resolved at our resolution.  We report the position
of the strongest feature in Table~\ref{table-masers}.  This feature is
offset even farther to the east and south than the 229~GHz masers,
with the caveat that it is difficult to determine the position of the
36~GHz maser accurately given the weakness of the maser and the
2\arcsec\ beamsize of the 36~GHz observations.

Several class~I masers appear to the south (36 and 44~GHz) and north
(44~GHz only) of the outflow.  Unlike in the outflow, the 36~GHz
masers in the south are usually not accompanied by cospatial 44~GHz
masers.  The northern and southern masers are generally redshifted and
blueshifted, respectively, compared to the bulk of the masers in the
outflow, although not all masers conform to this pattern.  The
southern masers are located closer to the sources DR21(OH)W,
DR21(OH)S, and the ridge of ammonia emission connecting them
(\citealt{mangum1992}; see also \citealt{sjouwerman2010}) than to the
outflow in DR21(OH), and therefore are unlikely to be associated with
the main source driving the outflow in DR21(OH).

\subsection{DR21W}

Despite the detection of only 5 maser features, DR21W contains the
brightest ($> 80$~Jy) 36~GHz maser in our sample.  This maser is the
only one for which we detect an antisymmetric Stokes V profile
(Figure~\ref{fig-zeeman}), which will be discussed further in
Section~\ref{magnetic}.  The bright masers near the origin in
Figure~\ref{fig-maps-n-w} are coincident with the 44~GHz masers
detected by \citet{kogan1998} to better than a synthesized beamwidth.
The velocities of the 36~GHz masers near the origin agree with the
velocities of the brightest nearby 44~GHz masers to within
0.2~km\,s$^{-1}$ or better.

\subsection{DR21N}

DR21N was first identified as a 36~GHz maser source by
\citet{pratap2008}.  We detect numerous 36~GHz masers in this source,
mostly in the range from $-6$ to $-4$~km\,s$^{-1}$, on the eastern
side of the distribution of maser emission
(Figure~\ref{fig-maps-n-w}).  Several redshifted masers in the range
$-1$ to $+5$~km\,s$^{-1}$ are found on the western side, along with
one highly blueshifted maser ($-14$~km\,s$^{-1}$).  All detected
229~GHz masers are found in close proximity to ($< 0\farcs5$ from) a
36~GHz maser.  In contrast, no 44~GHz maser is seen within
1\arcsec\ of a 36~GHz maser.

\section{Discussion}

\subsection{Circular Polarization}
\label{magnetic}

\begin{figure}
\resizebox{\hsize}{!}{\includegraphics{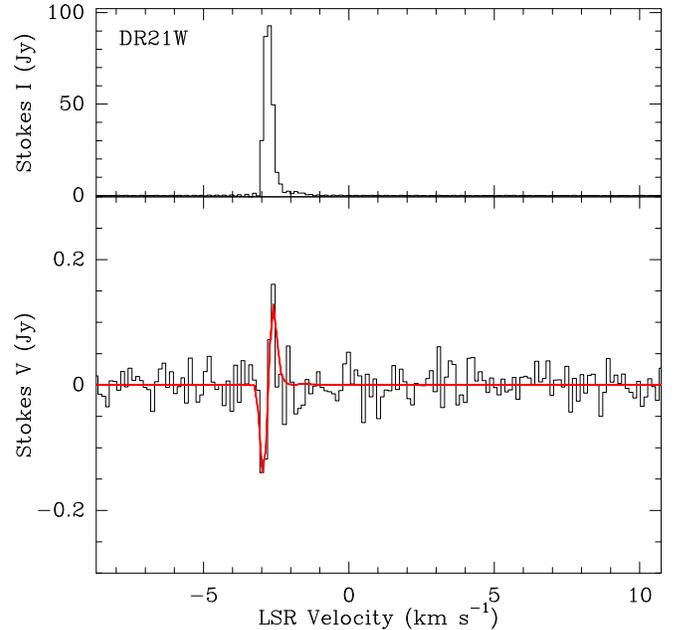}}
\caption{\emph{Top}: Stokes I spectrum of a $\sim 6\arcsec$
  rectangular region around the brightest maser feature in DR21W.
  \emph{Bottom}: Stokes V spectrum after a scaled version of Stokes I
  has been subtracted.  The thick red curve shows the best-fit value
  of the scaled derivative of Stokes I.  The detection of an
  ``S-curve'' in Stokes V, discussed in Section~\ref{magnetic}, is
  significant even if the spectra are Hanning smoothed to reduce the
  ringing evident in Stokes I.
\label{fig-zeeman}}
\end{figure}

As shown in Figure~\ref{fig-zeeman}, we detect an antisymmetric
``S-curve'' Stokes V signature in the spectrum of the brightest maser
in DR21W.  If interpreted as due to Zeeman splitting, the implied
line-of-sight magnetic field strength is $+58.1 \pm 6.2 ~ (1.7 ~
\mathrm{Hz}\,\mathrm{mG}^{-1}/z)$~mG, where $z$ represents the Zeeman
splitting coefficient \citep[assumed to be 1.7 Hz\,mG$^{-1}$ for the
  36~GHz transition by][]{sarma2009}.  The direction of this component
of the magnetic field would be oriented away from the observer.  The
full three-dimensional field strength would presumably be larger, as
Zeeman splitting measures only the line-of-sight component of the
magnetic field when the splitting between Zeeman components is much
less than the line width \citep{watson2001,vlemmings2008}.  Stokes V
S-curves were not detected in any of the other maser sources, with a
$3\,\sigma$ upper limit of $73 ~ (1.7 ~
\mathrm{Hz}\,\mathrm{mG}^{-1}/z)$~mG in the brightest remaining maser
in DR21(OH), although we note that a circular polarization fraction of
$\sim 0.2\%$ (i.e., the V/I seen in DR21W) would not produce a
detectable ($> 3\,\sigma$) Stokes V signal in DR21(OH) or in any of
the weaker masers.

However, there is significant uncertainty in the Zeeman splitting
coefficient, which calls into question the direct Zeeman
interpretation of methanol Stokes V S-curves.  \citet{jen1951}
obtained a laboratory estimate of the Land\'{e} $g$-factor for the
25~GHz $J_2 \rightarrow J_1$ series of transitions of methanol, but
the obtained value is an average over several different transitions
and, in any case, may not be appropriate for the 36~GHz transition of
methanol.  \citet{vlemmings2006} and \citet{vlemmings2008} used the
\citet{jen1951} $g$-factor to estimate the Zeeman splitting
coefficient for the 6.7~GHz transition, and \citet{sarma2009} followed
their method in deriving the Zeeman splitting coefficient at 36~GHz.
A more recent look suggests that these calculations may have
overestimated the Zeeman splitting coefficients by an order of
magnitude (W.~H.~T.\ Vlemmings 2010, private communication), which
would serve to \emph{increase} proportionally the putative magnetic
field strengths reported from methanol Zeeman splitting in this work
and others
\citep[e.g.,][]{vlemmings2008,surcis2009,sarma2009,sarma2010}.

Even if the \citet{sarma2009} Zeeman splitting coefficient for the
36~GHz transition is (approximately) correct, a line-of-sight magnetic
field strength of 58~mG is uncomfortably high.  Measurements of the
magnetic field strength are often used as a proxy for density.  Using
a scaling of $|B| \propto n^{0.47}$ from the densities and magnetic
fields of DR21 would give a number density around $10^{10}$~cm$^{-3}$
\citep{crutcher1999}, a value several orders of magnitude higher than
that though to be suitable for 36~GHz maser excitation
\citep[e.g.,][]{menten1991,johnston1992,sobolev2005}.  Scaling from
values typically found in OH masers instead \citep[a few milligauss,
  $n \approx 10^6$~cm$^{-3}$;][]{pavlakis1996,fish2005} results in
density estimates well in excess of $10^8$~cm$^{-3}$, which is still
substantially larger than theoretical estimates \citep[but see also
  Section 4.3 of][]{voronkov2005}.  The discrepancy between the
implied density and theoretical calculations appropriate for class~I
maser activity grows by two orders of magnitude if a smaller Zeeman
splitting factor is assumed (W.~H.~T.\ Vlemmings 2010, private
communication).

If the Zeeman interpretation is correct, one possible explanation for
the discrepancy is that the correlation between magnetic field
strength and density does not hold in (all) the environments of
class~I methanol masers, implying that the magnetic energy density is
much greater than the kinetic energy density \citep{sarma2009}.
Alternatively, it is possible that the high density implied by the
large magnetic field is correct, and the bright 36~GHz maser in DR21W
is produced via a pumping scheme not normally assumed for class~I
masers.  Typically, class~I masers are thought to be produced by
collisional excitation followed by spontaneous cascade down to lower
energy levels, with sink photons escaping or being absorbed by cold
dust.  This mechanism has an upper limit for density, above which the
inverted transition is thermalized by collisions.  One possible
pumping mechanism at higher densities involves collisional excitation
by a warm species and a collisional sink by a cooler species to
produce the population inversion \citep{strelnitskij1984}.  This
mechanism requires a two-temperature mixture of particles, such as
electrons (and/or ions) and neutrals, that may exist after the passage
of a shock front.  In principle, such a collisional-collisional pump
can operate at arbitrarily high density, as long as the temperature
difference between the two species is sustained.

However, it is more probable that the Stokes V signature we detect
does not measure the magnetic field directly.  \citet{wiebe1998}
proposed a mechanism by which changes in the magnetic field
orientation can convert linear polarization to circular polarization
along the amplification path in a maser.  The Stokes V signature
produced by this mechanism can produce an S-curve that can mimic the
effect of Zeeman splitting of a much larger magnetic field.  In the
case of circumstellar SiO masers that \citet{wiebe1998} consider,
their proposed mechanism could operate with a magnetic field as weak
as 0.1\% of the magnetic field strength implied from interpreting the
S-curve as purely due to Zeeman splitting.  The \citet{wiebe1998}
mechanism requires that the Zeeman splitting in units of frequency
exceed both the stimulated emission rate and the decay rate
appropriate for the masing transition, a condition likely to be
satisfied for class~I methanol masers \citep[see Section 4.2
  of][]{wiesemeyer2004}.

Interestingly, DR21W (this work) and M8E \citep{sarma2009}, the two
sources with detected Zeeman splitting in the 36~GHz line, are also
the two sources in which \citet{wiesemeyer2004} detected circular
polarization in the 133~GHz ($6_{-1} \rightarrow 5_0~E$) transition.
Future high-sensitivity observations at 133~GHz might permit detection
of an S-curve signature in Stokes V, thereby testing whether the
magnetic field implied from a pure Zeeman interpretation is consistent
with the values derived at 36~GHz.  (Since the 36~GHz and 133~GHz
methanol lines are both $J_{-1} \rightarrow (J-1)_0~E$ transitions,
the same Land\'{e} $g$-factor would be applicable to each of them.)
Careful laboratory measurements of the Zeeman splitting coefficients
appropriate for the brightest methanol maser transitions would also be
very helpful.

\subsection{A Class~I/Class~II Overlap}

Excitation models indicate that the physical conditions that are
thought to produce inversion in class~I transitions usually lead to
anti-inversion in class~II transitions and vice versa
\citep{cragg1992,slysh2002}.  In that regard, the existence of class~I
36 and 229~GHz masers at the same velocity and location as the
class~II 6.7~GHz maser site is surprising.  These masers are also
coincident with the brightest continuum emission near the outflow,
which may be an important clue to their origin.

There is precedent for overlaps between class~I and class~II methanol
masers.  \citet{voronkov2005} deduce spatial coincidences between the
6.7~GHz class~II and 25~GHz $J_2 \rightarrow J_1~E$ series of class~I
methanol masers in OMC-1, although the angular resolution of their
interferometric observations is much poorer than either the 229~GHz
data on which we report or the \citet{harvey-smith2008} 6.7~GHz data.
\citet{voronkov2005} model the 6.7/25~GHz overlap as arising from a
lower-temperature regime than is typically assumed for class~II
masers.  This pumping model requires an intermixed environment of gas
and dust at a lower temperature ($\sim 60$~K) than for traditional
class~II maser formation.  The strong 226~GHz continuum emission in
DR21(OH) likely marks such an environment that is rich in dust.  The
\citet{voronkov2005} model did not investigate excitation at 229~GHz,
but it does predict inversion at 36~GHz, which is also a $J_{-1}
\rightarrow (J-1)_0~E$ transition.  The model predicts that 44 and
95~GHz masers should \emph{not} appear at a low-temperature 6.7~GHz
maser site, since the former are $J_0 \rightarrow (J-1)_1~A^+$
transitions, while the latter is a $(J-1)_1 \rightarrow J_0~A^+$
transition.  Indeed, \citet{kogan1998}, \citet{kurtz2004}, and
\citet{araya2009} do not report 44~GHz masers at this location.

\subsection{Multitransition Comparison of Class~I Masers}

In each of the sources we observed, the 36~GHz masers divide into two
sets: those that are found to be cospatial with other class~I maser
transitions, and those that are spatially isolated.  We address each
of these two groups in turn.

In the outflow of DR21(OH) and in the cluster in the north of DR21W,
we note that the distribution of the 36~GHz masers is similar to that
of the 44, 84, 95, and 229~GHz masers.  Additionally, the brightest
masers in all of these transitions appear in the same location.  In
DR21(OH), the brightest masers in all class~I transitions occur in a
narrow velocity range ($v_\mathrm{LSR} = +0.3$ to $+0.5$~km\,s$^{-1}$)
at the western tip of the outflow.  The next-brightest masers in each
transition appear approximately 10\arcsec\ east of this position
(i.e., 10\arcsec\ west of the origin in Figure~\ref{fig-map-oh}).  It
is clear that these masers exist in a physical regime in which strong
maser emission is produced in a large number of $J_{-1} \rightarrow
(J-1)_0~E$ and $J_0 \rightarrow (J-1)_1~A^+$ transitions
simultaneously.  The existence of a nearby energetic source creating
an outflow likely shocks the surrounding molecular material, pumping
multiple transitions of methanol in the same location.

On the other hand, all sources contain regions in which 36~GHz masers
are not found to be coincident with other transitions.  For instance,
numerous masers in both the 36 and 44~GHz masers are found to the
south of the outflow in DR21(OH).  For the most part, the two
transitions are not seen to be spatially coincident.  Most of the
36~GHz masers and some of the 44~GHz masers in the south are seen near
the periphery of CS emission \citep{plambeck1990}.  The common feature
of the environment of these masers is the lack of a nearby energetic
source.  As in the Sagittarius~A region, it is possible that these
masers trace molecular density clumps that are shock-excited by core
collision or compression \citep{sjouwerman2010}.

The 229~GHz masers appear to align more closely with the 44~GHz masers
than with the 36~GHz masers in the DR21(OH) outflow.  While 229~GHz
masers are found to be coincident with 36~GHz masers on the western
side of the outflow, there is always an accompanying 44~GHz maser.  In
contrast, several 229~GHz masers appear coincident with 44~GHz masers
but without 36~GHz emission, especially on the eastern side of the
outflow.  However, we note several caveats about drawing conclusions
from these points.  First, the angular resolution of the observations
of both the 44~GHz and the 229~GHz masers is higher than that of the
36~GHz observations on which we report, and it is possible that more
sensitive, higher-resolution observations of the 36~GHz masers will
uncover other features at the sites of the 44/229~GHz overlaps.
Second, every detected 229~GHz maser in DR21N appears near a 36~GHz
maser site (Figure~\ref{fig-maps-n-w}).  Third, at present the three
DR21 sources on which we report are the only sources mapped in all
three of the 36, 44, and 229~GHz transitions, and the field of view
encompassed by the 229~GHz observations in DR21(OH) excludes nearly
all of the masers south of the outflow as well as the eastern tip of
the outflow itself.  This constitutes a rather small set of sources
from which to draw general conclusions about the properties of the 36,
44, and 229~GHz masers relative to each other.  More interferometric
observations of both the 36 and 229~GHz transitions will be required
to fully understand the phenomenology of either.

\section{Conclusions and Future Work}

We have interferometrically imaged three sources in the 36 and 229~GHz
class~I methanol maser lines.  We find numerous masers in both
transitions and a diversity of conditions among them.  Notably, we
identify the following three environments, each of which may correspond
to a different set of excitation conditions.

1. The outflow in DR21(OH) contains a large number of overlaps in the
36, 44, and 229~GHz transitions.  Strong emission is also seen at 84
and 95~GHz, although there is a lack of high-resolution observations
of these lines.  The brightest masers in all transitions appear at the
same velocity and in the same location at the western tip of the
outflow as traced by the masers.

2. The bright continuum source in DR21(OH) is associated with both the
class~II 6.7~GHz masers and class~I 36 and 229~GHz $E$-type masers,
but the class~I 44~GHz $A^+$-type masers are conspicuously absent.
This environment may be explained by the low-temperature intermixed
dust and gas model of \citet{voronkov2005}.

3. The 229~GHz transition produces detectable masers at a subset of
the 36~GHz maser sites in DR21N.  In both this source and in DR21(OH)
well south of the outflow, the 36 and 44~GHz masers have a similar
large-scale distribution but are rarely found to produce a maser at
the same site.

In addition, the brightest maser in DR21W produces an antisymmetric
Stokes V profile that implies a large magnetic field if interpreted as
being due to Zeeman splitting.  There is a very large uncertainty in
the Zeeman splitting coefficient appropriate for the 36~GHz
transition, but the implied density (assuming $|B| \propto n^\alpha,
\alpha \approx 0.5$) greatly exceeds the range over which class~I
masers are thought to form.  It is possible that the circular
polarization in this maser feature is produced by a much smaller
magnetic field whose orientation changes over the amplification path
of the maser \citep{wiebe1998}.  In any case, careful laboratory
measurement of Zeeman splitting coefficients appropriate for methanol
maser transitions is warranted given the increasing number of
Zeeman-like Stokes V signatures identified in methanol transitions
over the past few years.

In order to realize the goal of being able to identify the physical
conditions in a variety of star-forming regions by the presence or
absence of various methanol maser transitions, it will be necessary to
understand a few sources in greater detail.  Our results highlight the
need for both increased theoretical effort and more sensitive
observations of multiple class~I transitions at higher angular
resolution.  In particular, higher-resolution maps of the 84 and
95~GHz transitions as well as higher-frequency maser lines may be both
enlightening and timely in the advent of the ALMA era.  A further
survey of sources in the 36 and 229~GHz transitions, when combined
with 44~GHz maps in the literature, may also be helpful in determining
the range of possible methanol excitation conditions in
nature.

\acknowledgments

T.~C.~M.\ acknowledges support from the National Science Foundation's
Research Experiences for Undergraduates program.  The National Radio
Astronomy Observatory is a facility of the National Science Foundation
(NSF) operated under cooperative agreement by Associated Universities,
Inc.  The Submillimeter Array is a joint project between the
Smithsonian Astrophysical Observatory and the Academia Sinica
Institute of Astronomy and Astrophysics and is funded by the
Smithsonian Institution and the Academia Sinica.  We thank
W.~H.~T.\ Vlemmings and A.\ Sarma for helpful discussions regarding
methanol Zeeman splitting coefficients.

{\it Facilities:} \facility{EVLA ()}, \facility{SMA ()}



\begin{thebibliography}{}

\bibitem[Araya et al.(2009)]{araya2009} Araya, E.~D., Kurtz, S.,
  Hofner, P., \& Linz, J.\ 2009, \apj, 698, 1321

\bibitem[Batrla \& Menten(1988)]{batrla1988} Batrla, W., \& Menten,
  K.~M.\ 1988, \apj, 329, L117

\bibitem[Berulis et al.(1990)]{berulis1990} Berulis, I.~I., Kalenskii,
  S.~V., \& Logvinenko, S.~V.\ 1990, Sov.\ Astron.\ Lett., 16, 179

\bibitem[Breen et al.(2010)]{breen2010} Breen, S.~L., Ellingsen,
  S.~P., Caswell, J.~L., \& Lewis, B.~E.\ 2010, \mnras, 401, 2219

\bibitem[Cragg et al.(1992)]{cragg1992} Cragg, D.~M., Johns, K.~P.,
  Godfrey, P.~D., \& Brown, R.~D.\ 1992, \mnras, 259, 203

\bibitem[Crutcher(1999)]{crutcher1999} Crutcher, R.~M.\ 1999, \apj,
  520, 706

\bibitem[Ellingsen(2006)]{ellingsen2006} Ellingsen, S.~P.\ 2006, \apj,
  638, 241

\bibitem[Fish et al.(2005)]{fish2005} Fish, V.~L., Reid, M.~J., Argon,
  A.~L., \& Zheng, X.-W.\ 2005, \apjs, 160, 220

\bibitem[Harvey-Smith et al.(2008)]{harvey-smith2008} Harvey-Smith,
  L., Soria-Ruiz, R., Duarte-Cabral, A., \& Cohen, R.~J.\ 2008,
  \mnras, 384, 719

\bibitem[Haschick \& Baan(1989)]{haschick1989} Haschick, A.~D., \&
  Baan, W.~A.\ 1989, \apj, 339, 949

\bibitem[Jen(1951)]{jen1951} Jen, C.~K.\ 1951, Physical Review, 81,
  197

\bibitem[Johnston et al.(1992)]{johnston1992} Johnston, K.~J., Gaume,
  R., Stolovy, S., Wilson, T.~L., Walmsley, C.~M., \& Menten,
  K.~M.\ 1992, \apj, 385, 232

\bibitem[Kalenskii et al.(1994)]{kalenskii1994} Kalenskii, S.~V.,
  Liljestr\"{o}m, T., Val'tts, I.~E., Vasil'kov, V.~I., Slysh, V.~I.,
  \& Urpo, S.\ 1994, \aaps, 103, 129

\bibitem[Kogan \& Slysh(1998)]{kogan1998} Kogan, L., \& Slysh,
  V.\ 1998, \apj, 497, 800

\bibitem[Kurtz et al.(2004)]{kurtz2004} Kurtz, S., Hofner, P., \&
  \'{A}lvarez, C.~V.\ 2004, \apjs, 155, 149

\bibitem[Liechti \& Wilson(1996)]{liechti1996} Liechti, S., \& Wilson,
  T.~L.\ 1996, \aap, 314, 615

\bibitem[Mangum et al.(1992)]{mangum1992} Mangum, J.~G., Wootten, A.,
  \& Mundy, L.~G.\ 1992, \apj, 388, 467

\bibitem[Mehringer \& Menten(1997)]{mehringer1997} Mehringer, D.~M.,
  \& Menten, K.~M.\ 1997, \apj, 474, 346

\bibitem[Menten(1991)]{menten1991} Menten, K.~M.\ 1991, in ASP
  Conf.\ Ser.\ 16, Atoms, Ions and Molecules: New Results in Spectral
  Line Astrophysics, ed.\ A.~D.~Haschick \& P.~T.~P.~Ho (San
  Francisco: ASP), 119

\bibitem[M\"{u}ller et al.(2004)]{muller2004} M\"{u}ller, H.~S.~P.,
  Menten, K.~M., \& M\"{a}der, H.\ 2004, \aap, 428, 1019

\bibitem[Pavlakis \& Kylafis(1996)]{pavlakis1996} Pavlakis, K.~G., \&
  Kylafis, N.~D.\ 1996, \apj, 467, 309

\bibitem[Plambeck \& Menten(1990)]{plambeck1990} Plambeck. R.~L., \&
  Menten, K.~M.\ 1990, \apj, 364, 555

\bibitem[Pratap et al.(2008)]{pratap2008} Pratap, P., Shute, P.~A.,
  Keane, T.~C., Battersby, C., \& Sterling, S.\ 2008, \apj, 135, 1718

\bibitem[Sarma \& Momjian(2009)]{sarma2009} Sarma, A.~P., \& Momjian,
  E.\ 2009, \apj, 705, L176

\bibitem[Sarma \& Momjian(2010)]{sarma2010} Sarma, A.~P., \& Momjian,
  E.\ 2010, in International SKA Forum 2010 Science Meeting,
  PoS(ISKAF2010), 074

\bibitem[Sjouwerman et al.(2010)]{sjouwerman2010} Sjouwerman, L.~O.,
  Pihlstr\"{o}m, Y.~M., \& Fish, V.~L.\ 2010, \apj, 710, L111

\bibitem[Slysh et al.(2002)]{slysh2002} Slysh, V.~I., Kalenski\u{\i},
  S.~V., \& Val'tts, I.~E.\ 2002, Astronomy Reports, 46, 49

\bibitem[Slysh et al.(1994)]{slysh1994} Slush, V.~I., Kalenskii,
  S.~V., Val'tts, I.~E., \& Otrupcek, R.\ 1994, \mnras, 268, 464

\bibitem[Sobolev(1993)]{sobolev1993} Sobolev, A.~M.\ 1993,
  Astron.\ Lett., 19, 293

\bibitem[Sobolev et al.(2005)]{sobolev2005} Sobolev, A.~M.,
  Ostrovskii, A.~B., Kirsanova, M.~S., Shelemei, O.~V., Voronkov,
  M.~A., \& Malyshev, A.~V.\ 2005, in IAU Symp.\ 227, Massive Star
  Birth: A Crossroads of Astrophysics, ed.\ R.~Cesaroni, M.~Felli,
  E.~Churchwell, \& M.~Walmsley (Cambridge: Cambridge Univ.\ Press),
  174

\bibitem[Strelnitskij(1984)]{strelnitskij1984} Strelnitskij,
  V.~S.\ 1984, \mnras, 207, 339

\bibitem[Surcis et al.(2009)]{surcis2009} Surcis, G., Vlemmings,
  W.~H.~T., Dodson, R., \& van Langevelde, H.~J.\ 2009, \aap, 506, 757

\bibitem[Vlemmings(2008)]{vlemmings2008} Vlemmings, W.~H.~T.\ 2008,
  \aap, 484, 773

\bibitem[Vlemmings et al.(2006)]{vlemmings2006} Vlemmings, W.~H.~T.,
  Harvey-Smith, L., \& Cohen, R.~J.\ 2006, \mnras, 371, L26

\bibitem[Voronkov et al.(2006)]{voronkov2006} Voronkov, M.~A., Brooks,
  K.~J., Sobolev, A.~M., Ellingsen, S.~P., Ostrovskii, A.~B., \&
  Caswell, J.~L.\ 2006, \mnras, 373, 411

\bibitem[Voronkov et al.(2010a)]{voronkov2010} Voronkov, M.~A.,
  Caswell, J.~L., Britton, T.~R., Green, J.~A., Sobolev, A.~M., \&
  Ellingsen, S.~P.\ 2010a, \mnras, in press

\bibitem[Voronkov et al.(2010b)]{voronkov2010-9.9} Voronkov, M.~A.,
  Caswell, J.~L., Ellingsen, S.~P., \& Sobolev, A.~M.\ 2010b, \mnras,
  405, 2471

\bibitem[Voronkov et al.(2005)]{voronkov2005} Voronkov, M.~A.,
  Sobolev, A.~M., Ellingsen, S.~P., \& Ostrovskii, A.~B.\ 2005,
  \mnras, 362, 995

\bibitem[Watson \& Wyld(2001)]{watson2001} Watson, W.~D., \& Wyld,
  H.~W.\ 2001, \apj, 558, L55

\bibitem[Wiebe \& Watson(1998)]{wiebe1998} Wiebe, D.~S., \& Watson,
  W.~D.\ 1998, \apj, 503, L71

\bibitem[Wiesemeyer et al.(2004)]{wiesemeyer2004} Wiesemeyer, H.,
  Thum, C., \& Walmsley, C.~M.\ 2004, \aap, 428, 479

\end{thebibliography}
\end{document}